**Spin-valve effects in point contacts to exchange biased $Co_{40}Fe_{40}B_{20}$ films**

**O. P. Balkashin[1], V. V. Fisun[1], L. Yu. Triputen[1], S. Andersson[2], V. Korenivski[2], Yu. G. Naidyuk [1]**

[1] B.Verkin Institute for Low Temperature Physics and Engineering, National Academy of Sciences of Ukraine, 47 Lenin Ave., 61103, Kharkiv, Ukraine

[2]Nanostructure Physics, Royal Institute of Technology, Stockholm, SE-10691 Sweden

Abstract.

Nonlinear current-voltage characteristics and magnetoresistance of point contacts between a normal metal (N) and films of amorphous ferromagnet (F) $Co_{40}Fe_{40}B_{20}$ of different thickness, exchange-biased by antiferromagnetic $Mn_{80}Ir_{20}$ are studied. A surface spin valve effect in the conductance of such F–N contacts is observed. The effect of exchange bias is found to be inversely proportional to the $Co_{40}Fe_{40}B_{20}$ film thickness. This behavior as well as other magneto-transport effects we observe on single exchange-pinned ferromagnetic films are similar in nature to those found in conventional three-layer spin-valves.

Interest in the study of layered systems ferromagnet/antiferomagnet (F/AF) is due to some open fundamental questions as to the mechanisms of exchange bias as well as applications of such structures in spintronic devices [1] exploiting the giant magnetoresistance effect [2]. Exchange bias was first discovered in micro particles of cobalt oxide CoO [ 3], manifested as a field-shift of the hysteresis loop with respect to H=0. Nowadays, this effect is widely used for pinning the magnetization of one of the layers of a spin valve (SV ) [4], which typically consists of two ferromagnetic films separated by a normal metal, $F_1/N/F_2$. In the antiparallel configuration of the magnetization of the $F_{1,2}$ layers, the resistance of the SV is high, and is lowered on switching into the parallel configuration. This change in the resistance in response to an external field gives the SV its key functionality as a field sensor. For pinning the direction of the magnetization of the ferromagnetic layer the following procedure is used: at a temperature above the Neel temperature $T_N$ of the antiferromagnet (AF), but below the Curie temperature $T_C$ of the ferromagnet, the AF is magnetically disordered while F is magnetically ordered. As the temperature is decreased below $T_N$, with a suitably high magnetic field applied to the sample (to achieve saturation of F), the AF becomes magnetically ordered, with its interface spins aligned by the ferromagnet. This preferred spin direction in the AF, induced by the field-cooling, acts as magnetic bias when the F-magnetization is switched by a reversing magnetic field. The AF-bias results in a shift of the hysteresis loop of F proportional to the interfacial exchange force, with its quantifying parameter, the so-called exchange-bias field $H_{EX}$, defined as

$$H_{EX} = - (H_L + H_R)/2. \quad (1)$$

Here $H_L$ и $H_R$ are the magnetic field values at which the magnetization reversal in F occurs.

The growing interest in magnetic nanostructures based on multicomponent amorphous alloys is due to the fact that the realization of the effect of spin transfer torque in such systems [5] requires smaller switching currents as compared with single-element ferromagnets [6]. In this regard, amorphous alloy CoFeB is a promising candidate for applications in devices based on the giant magnetoresistance effect due to its high electrical resistance and low magnetic anisotropy [7,8 ], which are beneficial to functional parameters of spintronic nanostructures. In particular, a tunnel junction CoFeB/MgO/CoFeB with epitaxial films of magnesium oxide has been demonstrated to have giant (up to one thousand percent) magnetoresistance, between its parallel and antiparallel magnetic configurations [9], which is due to due to a nearly full spin asymmetry of the electron transport through the CoFeB/MgO interface.

The key magnetic properties for a variety of spintronic devices are the coercive field of the magnetic material, its saturation magnetization, weak magnetostriction, all present in CoFeB, which is why the material is widely used in research and industrial applications. The alloy is amorphous and therefore can optimally grow on almost any substrate, avoiding lattice mismatch between the under- and over-layers and minimizing other defects in the stack.

In our previous papers [10–13] it was shown that in nonmagnetic-ferromagnetic single-interface tip-surface point contacts, with the ferromagnet being polycrystalline Co, the magneto-transport exhibits many of the characteristic of traditional pillar-like $F_1$/N/$F_2$ spin-valves (with two magnetic interfaces, having parallel and antiparallel spin configuration). Our analysis of the experiments resulted in a spin-transport model for a single F/N interface in the point contact region [11, 13]. The model proposes that an atomically thin surface layer of spins in cobalt is magnetically rotated with respect to the spins in the bulk, forming an atomically thin domain wall. It is well known that in narrow contacts between two ferromagnetic electrodes, the width of a magnetic domain wall can be sufficiently small, commensurate with the size of the narrowing [14]. Such a thin domain wall allows to have orientation of the surface layer magnetization different from the bulk. As a result, in a simple point contact, with a single ferromagnetic film, two magnetic subsystem are realized – the bulk of the film and its surface layer with different anisotropy and coercivity values, forming a spatial structure of a “surface spin-valve” [11].

In [12], analyzing the dependence of the magnetoresistance of the surface SV based on an exchange biased Co/FeMn film, we found that the exchange bias shifts not only the inner transitions of the magnetoresistance loop, corresponding to the magnetization reversal of the bulk of the Co film, but also the external transitions coming from the reversal of the surface layer. Thus, despite the weak coupling between the surface layer and the bulk of the film, the coupling is sufficient to offset the switching of the surface layer.

The aim of this work was to determine the effect of the thickness of the ferromagnetic layer on the magnitude of the exchange bias acting on the bulk of the film and the surface layer, using point contacts based on amorphous $Co_{40}Fe_{40}B_{20}$. It is known from the literature [15,16] that increasing the thickness of the ferromagnetic film should result in diminishing the effect of the exchange bias, functionally according to ~1/*t*, where *t* is the thickness of the ferromagnet.

For this study AF/F exchange biased bi-layers were used, with AF being $Mn_{80}Ir_{20}$ and F - $Co_{40}Fe_{40}B_{20}$. The exchange bias was set in the standard way: by slowly cooling the sample in a magnetic field of 350 Oe, from temperature higher than the blocking temperature of the AF ($T_b$ ~ 513 K for $Mn_{80}Ir_{20}$) to room temperature.

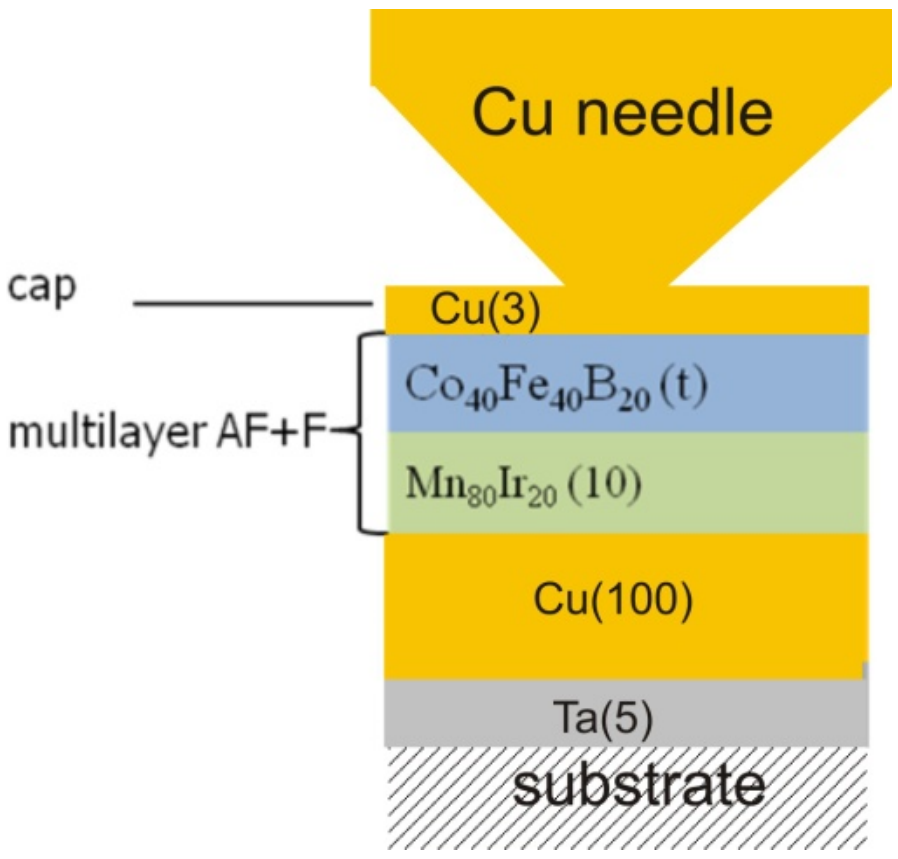

**Fig.1.** Schematic of a tip-surface N/F/AF point contact. The layer thicknesses are shown in parentheses in nanometers.

The studied samples were multilayer films deposited on oxidized Si substrates and schematically are shown in Fig.1. A seed layer of Ta was first deposited on to the substrate, followed by copper and antiferromagnetic $Mn_{80}Ir_{20}$, for exchange-biasing the $Co_{40}Fe_{40}B_{20}$ overlayer. The structure was capped with a 3 nm thick copper layer to protect the ferromagnet against oxidation. All layers were deposited in-situ, without breaking the vacuum, for ensuring high-quality interfaces. The tip was made out of a thin copper wire, sharpened into to a needle, first mechanically, then by chemical etching in $HNO_3$.

The measurements of the differential resistance *dV/dI(V)* of the point contacts as a function of the applied voltage bias were made using the traditional method of synchronous detection of the amplitude of a modulated ac-signal. The resistance of the studied point contacts was in the range 7–30 Ω, and their radii, estimated using Maxwell's formula [17], were 35–10 nm. This estimate used the typical value of the resistivity of $Co_{78}Fe_{11}B_{11}$ of 100 μΩcm [18]. External magnetic field was applied parallel to the film plane. All measurements were made at 4.2 K.

Figure 2 shows the dependence of the differential resistance *dV/dI(V)* in a magnetic field, where a peak is seen at the negative bias polarity. The peak position shifts toward higher currents with increasing magnetic field. A similar effect, the “dynamic SV effect”, was observed in point contacts based on cobalt [10,19], and was attributed to the excitation of a stationary precession of the magnetization vector due to a transfer of the spin angular momentum of the transport electrons to the angular momentum of the ferromagnetic lattice. Thus, the dynamic SV effect is also present in amorphous magnetic films, in our case $Co_{40}Fe_{40}B_{20}$. Notice that coefficient k (see inset in Figure 2) has the value corresponding to the lower bound for the same coefficient measured on crystalline cobalt point contacts [19]. Hence we can conclude that the excitation of the single-interface SV effects in $Co_{40}Fe_{40}B_{20}$ requires lower driving current densities than in the case of pure cobalt.

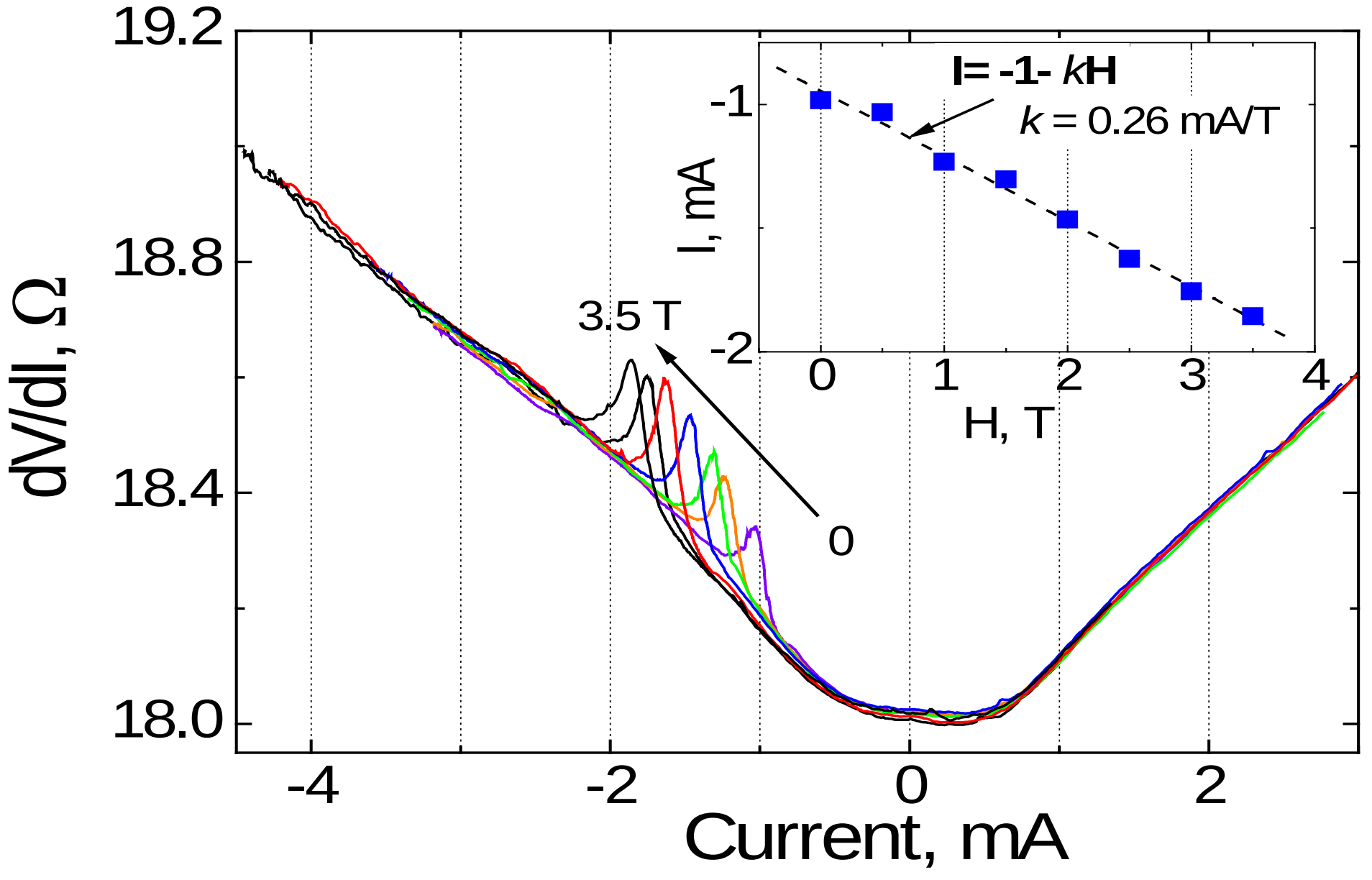

Fig.2. Spin-transfer peak in *dV/dI (I)* for a typical point contact of $Co_{40}Fe_{40}B_{20}$ (t = 20 nm) – Cu, for applied magnetic field in the range 0 to 3.5 T. The negative bias corresponds to the current direction flow from the film into the tip (electrons from the tip into the magnetic film). The inset shows the peak position as a function of the magnetic field magnitude.

Figure 3 shows the dependence of *dV/dI(V)* and magnetoresistance *dV/I(V=0,H)* of point contacts to $Co_{40}Fe_{40}B_{20}$ films of various thickness (20, 9, 6, 3 nm). It is known for spin valves [5] that sweeping the magnetic field results in magnetoresistance that has the form of two rectangle-like meanders or loops. Fig. 3 (b, d, f, h) show our measured magnetoresistance as two such loops, most clearly seen for the CoFeB film thickness of 20 nm. The switching observed in the magnetoresistance loop is between two stable spin configurations in the point contact core, and by analogy with traditional three-layer spin valves, between the parallel and antiparallel of the single-interface spin valve in our case. Comparing the *dV/dI(V)* of Figure 3a and 3b, where a clear rectangle-shaped hysteresis is observed, one can notice that the magnitude of the resistance change in *dV/dI(V)* and *dV/dI(H)* is essentially the same. This indicates that the effect of the electron spin-transfer torque and the effect of the magnetic field is on the same magnetic system, in the core of the point contact as that is what determines the measured resistance.

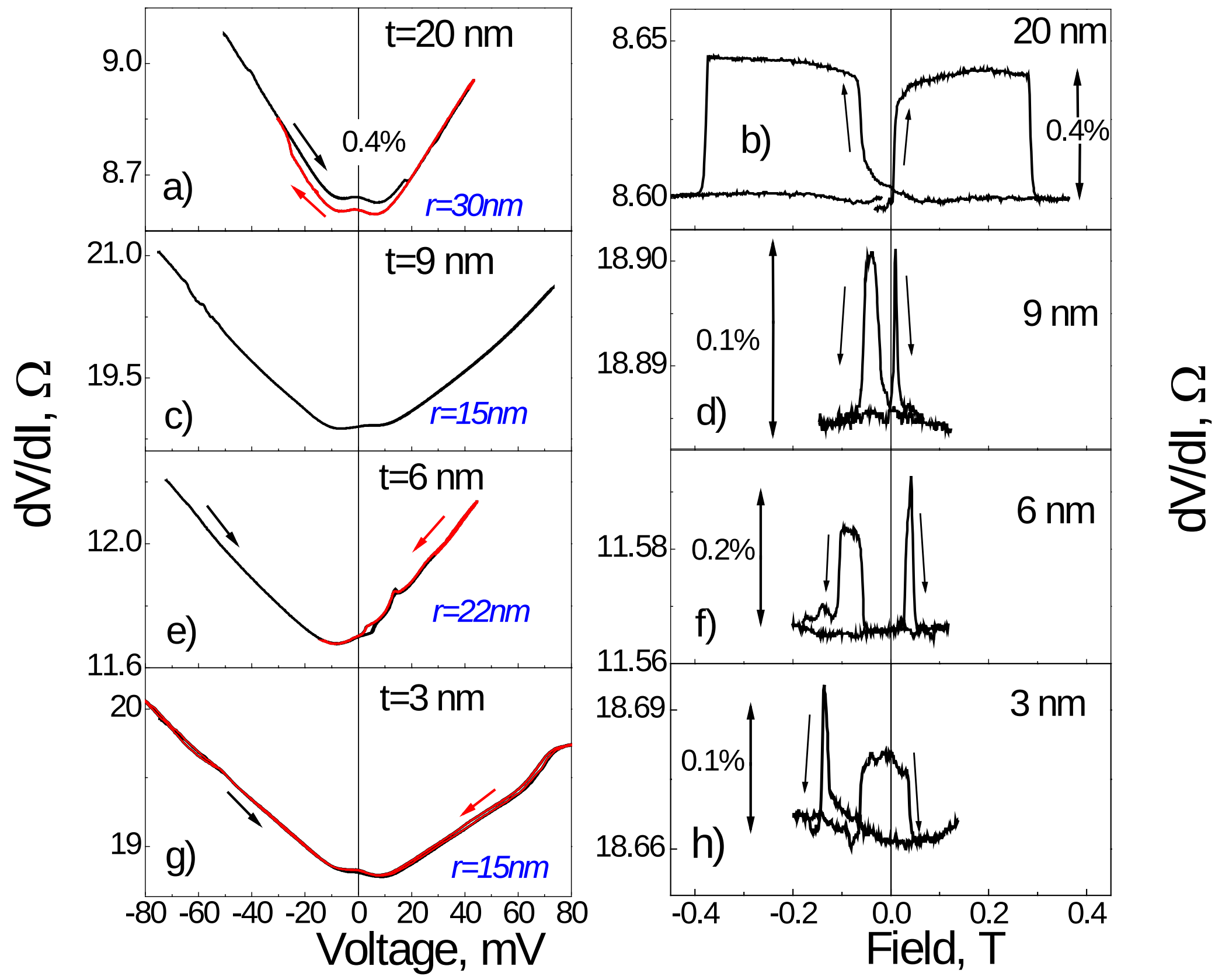

Fig. 3. *dV/dI(V)* and the corresponding magnetoresistance *dV/dI(H)* for contacts $Co_{40}Fe_{40}B_{20}$ (t) – Cu (t = 3, 6, 9, 20 nm). Arrows indicate the direction of the respective sweep, from left to right and from right to left. Text legends show the film thickness, t, value of the radii of the contacts, r, estimated using the Maxwell formula [16], and the scale of the magnetoresistance in %.

The lack of apparent hysteresis in the *dV/dI(V)* for smaller thicknesses (t = 3, 6, 9 nm) may be due to different values of the exchange-pinning strength at the surface layer (switching layer). For example, the curve in Fig.3g does show hysteresis with, however, significantly washed out transitions between the low- and high-resistance states. It is possible that in this particular case, the coercivity of the magnetic surface layer is large such that the spin transfer from the electron current is not sufficient for reversing the layer's magnetization. Additional to the washed out switching in thinner films, the corresponding change in the resistance is about 0.1%, which is several times smaller than that for 20 nm. The rather sharp peaks of the respective R-H curves are similar to the dependence characteristic of anisotropic magnetoresistance. Indeed, as shown in [20], the anisotropic magnetoresistance has a typical value of 0.15% for the composition $Co_{73.8}Fe_{16.2}B_{10}$. It is therefore possible that our thinner films exhibit predominantly anisotropic magnetoresistance. It may be informative to note that for the thinner films the typical point contact size exceeds the film thickness, so the ferromagnet occupies a smaller portion of the contact core and may contribute less to the measured resistance.

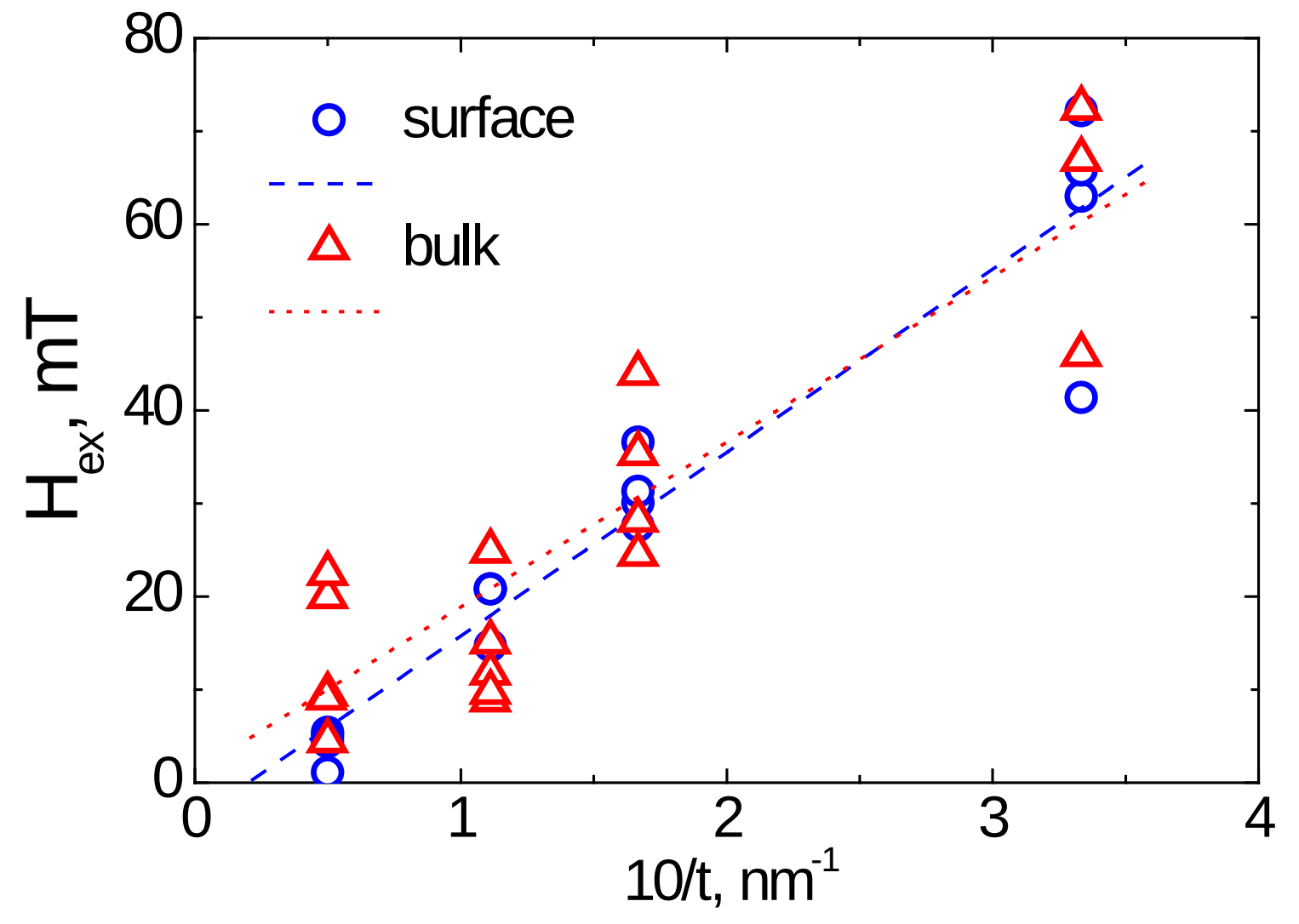

Fig. 4. Exchange-bias field $H_{EX}$ for $Co_{40}Fe_{40}B_{20}/Mn_{80}Ir_{20}$ versus the thickness of the ferromagnetic film: circles correspond to the surface layer of the film, triangles – to the bulk of the film. Straight lines: linear fits to the corresponding data points.

Figure 4 shows the dependence of the exchange bias field, calculated using equation (1) for several contacts, versus the inverse of the thickness of $Co_{40}Fe_{40}B_{20}$. The values of the characteristic fields of the magnetization reversal, $H_B$ (for bulk) and $H_S$ (surface) are determined from the R-H curves as positions of the switching boundaries (inner and outer) on the H-axis (fig.3b, d, f, h), for both directions of the field.

As can be seen from Fig. 4, with decreasing thickness of the ferromagnet, the exchange field increases. This is consistent with the literature data [15,16] and is due to the fact that the thinner ferromagnet is more strongly "pinned" by the antiferromagnet and, as a consequence, it requires larger values of the applied field for the magnetization reversal. In addition, the field strength must be sufficient to overcome the self-pinning by the surface mentioned above. It was also found that the surface layer also is affected by the exchange bias, because the outer switching field is asymmetric with respect to H = 0. Apparently, the magnetic subsystem of the film and the surface layer have a magnetic coupling. Similar field offsets were observed for exchange-biased films of polycrystalline cobalt [12].

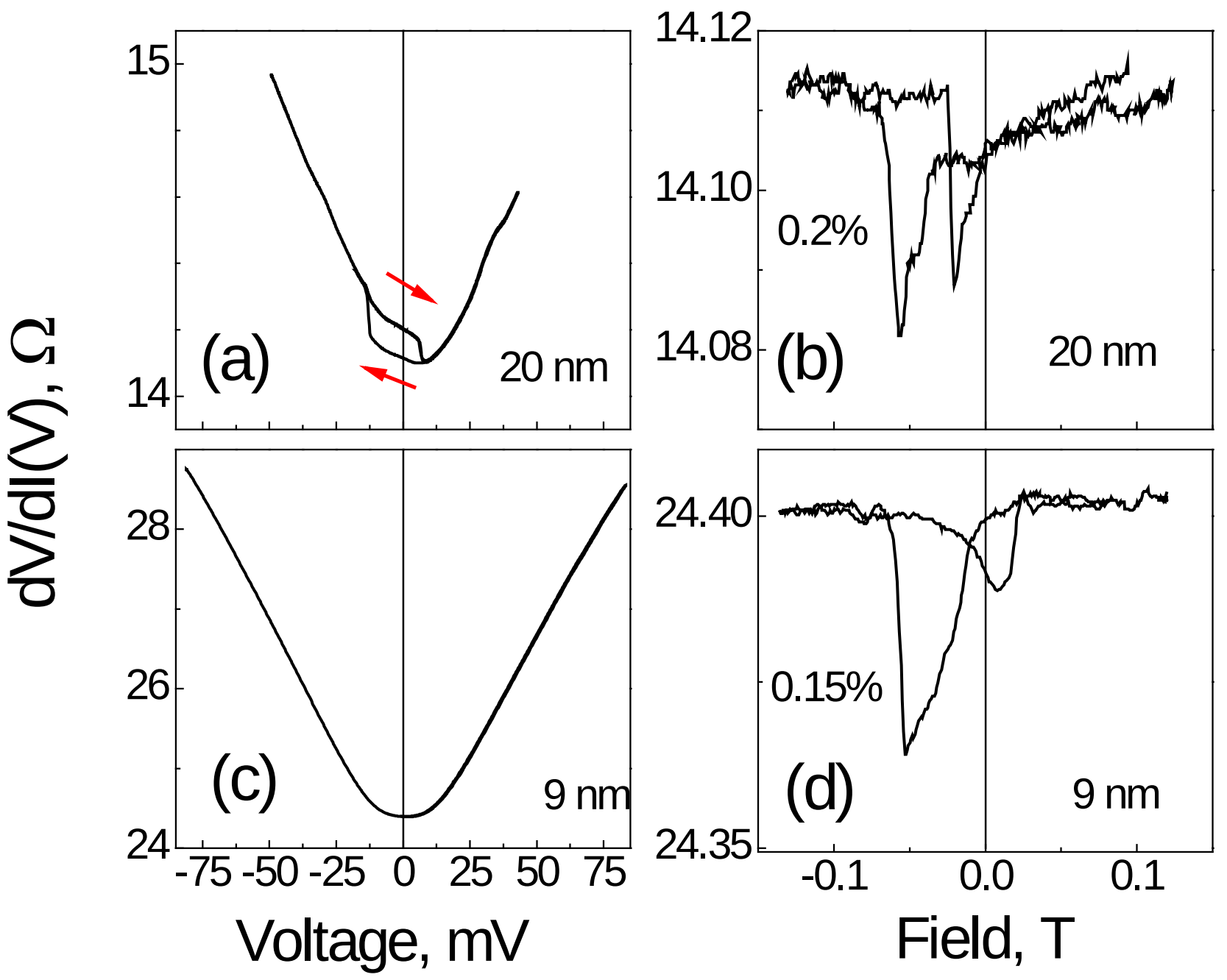

Figure 5. *dV/dI(V)* and the corresponding magnetoresistance *dV/dI(H)* for two contacts $Co_{40}Fe_{40}B_{20}$ (t)–Cu (t=9, 20 nm), exhibiting inverse magnetoresistance. The scale in percent near *dV/dI(H)* shows the magnitude of the MR.

In some cases, *dV/dI(H)* showed inverse magnetoresistance (high resistance at high field), illustrated in Fig. 5, in contrast to the more common positive MR (low resistance at high field). Figure 6 shows schematically a transport model for anisotropic magnetoresistance, where MR inversion is possible if the current is predominantly in the plane of the film. We also note that a qualitatively similar picture arises in the case of perpendicular magnetic anisotropy, i.e., when the favored direction of the magnetization in the ferromagnetic film is orthogonal to its plane. Perpendicular magnetic anisotropy is usually characteristic of very thin ferromagnetic films due to the effect of the interfaces, however, in [21] it has been observed in $Co_{40}Fe_{40}B_{20}$ films with thicknesses up to 172 nm. This was attributed to the presence of special (300) texture in the film. In our case, the cause of the perpendicular anisotropy can potentially be the mechanical stress produced by the tip in the point contact core region [23]. Thus, the observed inversion of the magnetoresistance can be due to a significant in-plane current flow and/or perpendicular magnetic anisotropy in the contact region. It may be informative to mention that no MR inversion was observed in point contacts to exchange-biased cobalt films, $Co/Fe_{50}Mn_{50}$ [12]. One more potential reason for the inversed MR may be a change of the sign of the spin asymmetry coefficient β of the current from positive to negative, which is observed for Cr, V, or Mn impurities in ferromagnets (see Table 1 and Figure 10 of [23]). In our case, the latter is possible due to the presence of a Mn-rich layer in the stack.

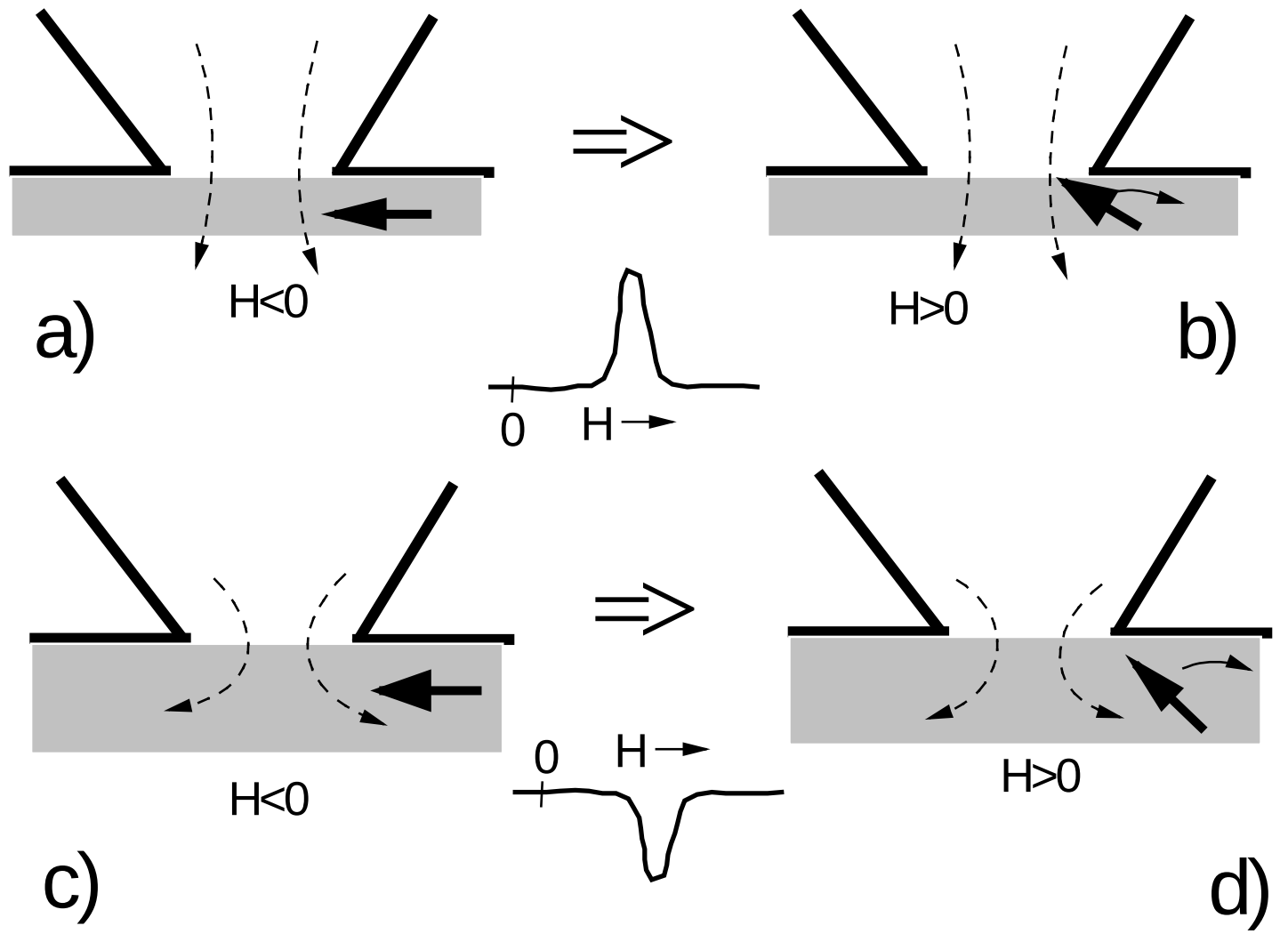

Figure 6. Schematic of a N/F tip-surface point contact. The ferromagnetic film is shown as a gray rectangle. Bold arrows show the direction of the magnetization **M**, and the dashed arrows represent the current flow lines in the contact. It is known that anisotropic magnetoresistance is smaller when the current flows perpendicular to **M** [22]. Two graphs in the middle illustrate the anisotropic magnetoresistance behavior with increasing magnetic field: a) and b) is the case of the current flow predominantly perpendicular to the film plane, and c) and d) is the case of the dominating contribution from the current component in the film plane. It is assumed that **M** has an out of plane component in both cases.

## Conclusion

We have studied the differential resistance *dV/dI(V)* and magnetoresistance *dV/dI (H, V=0)* of point contacts based on films of amorphous ferromagnetic $Co_{40}Fe_{40}B_{20}$ of different thickness, exchange-biased by antiferromagnetic $Mn_{80}Ir_{20}$. We have observed a single-interface spin-valve effect, characteristic of the conventional three-layer nanopillars $F_1$/N/$F_2$. The strength of the exchange bias in the bulk as well as at the surface of CoFeB is inversely proportional to the thickness of the ferromagnetic film, which demonstrates that point contacts to multicomponent amorphous single-layer ferromagnetic films can behave similar to the traditional multi-layer spin valves. We additionally observe an unusual magnetoresistance inversion in the studied amorphous exchange-biased system and discuss its possible origins.

## Acknowledgements

Funding by the National Academy of Sciences of Ukraine under project 26/14-H (NANO) is gratefully acknowledged.